\documentclass[11pt,superscriptaddress,aps,prd,preprint,showpacs]{revtex4}

\usepackage[dvips]{graphicx}
\usepackage{amsfonts}
\usepackage{slashed}
\usepackage[utf8x]{inputenc}
\usepackage{amsmath}
\usepackage{hyperref}

\begin{document}

%\title{CPT-violating Horava-Lifshitz $z=3$ electrodynamics and generation of the Chern-Simons term}
\title{CPT-violating $z=3$ Horava-Lifshitz QED and generation of the Carroll-Field-Jackiw term}

\author{T. Mariz}
\affiliation{Instituto de F\'{\i}sica, Universidade Federal de Alagoas,\\ 57072-900, Macei\'o, Alagoas, Brazil}
\email{tmariz,rmartinez@fis.ufal.br}

\author{R. Martinez}
\affiliation{Instituto de F\'{\i}sica, Universidade Federal de Alagoas,\\ 57072-900, Macei\'o, Alagoas, Brazil}
\email{tmariz,rmartinez@fis.ufal.br}

\author{J. R. Nascimento}
\affiliation{Departamento de F\'{\i}sica, Universidade Federal da Para\'{\i}ba,\\
 Caixa Postal 5008, 58051-970, Jo\~ao Pessoa, Para\'{\i}ba, Brazil}
\email{jroberto,petrov@fisica.ufpb.br}

\author{A. Yu. Petrov}
\affiliation{Departamento de F\'{\i}sica, Universidade Federal da Para\'{\i}ba,\\
 Caixa Postal 5008, 58051-970, Jo\~ao Pessoa, Para\'{\i}ba, Brazil}
\email{jroberto,petrov@fisica.ufpb.br}

\begin{abstract}
We study the $z=3$ Horava-Lifshitz QED with a CPT-breaking term, characterized by the axial vector $b_{\mu}$, and perform the one-loop calculations. Explicitly, we demonstrate that just as in the usual Lorentz-breaking QED, in our case, the Carroll-Field-Jackiw term is finite but ambiguous.
\end{abstract}


\maketitle

The theories with space-time anisotropy, also called the Horava-Lifshitz (HL) theories, are intensively studied now. Being introduced many years ago within the context of phase transitions \cite{Lifshitz}, they acquired great attention after publishing of the seminal paper \cite{Horava:2009uw}, where the space-time anisotropy has been implemented within the gravity context in order to construct power-counting renormalizable ghost-free gravity model. Moreover, HL modifications not only of gravity but also of other theories began to be studied, especially, various version of scalar and spinor QED, with various values of the critical exponent $z$, mostly $z=2$ and $z=3$. Among the most important results, one should emphasize the explicit calculation of the one-loop effective potential \cite{EP} and the one-loop counterterms \cite{CT}.

One of the reasons for the interest in HL-like theories is based on the idea of emergent dynamics. Following this concept, the known field theory models are generated as quantum corrections in some fundamental theory involving only spinors. Within the HL context, this methodology has been applied to the Gross-Neveu model, where the effective dynamics of a (pseudo)scalar field arises \cite{Lima}. Moreover, within this methodology, dynamics not only for a scalar but also for a vector field has been generated \cite{Ruben}.

Clearly, more studies can be performed within the HL context. In particular, it is interesting to study HL-like theories where the CPT symmetry is broken, which can be done through the coupling of fermions to a constant axial vector, similar to the scheme used in a Lorentz-breaking QED to generate the Carroll-Field-Jackiw (CFJ) term. For the first time, this generation has been performed in \cite{JK} (for a general review on obtaining this term, see \cite{ourrev} and references therein), so, as a result of integration over fermions, in the first order of expansion of the effective action in this constant axial vector, the CFJ term arises being finite and ambiguous. In this paper, we use this approach to study the possibility of arising the one-loop CFJ term in the $z=3$ HL QED (we note that in $z=2$ HL QED the CFJ term does not arise \cite{TMP}), that is, our aim consists in the study of the low-energy effective dynamics in this theory. The importance of our study is confirmed by the fact that, up to now, the breaking of the CPT symmetry, usually implemented with the introduction of some constant axial vector and/or $\gamma_5$ matrix, never has been performed in HL-like theories.

We start with introducing the CPT-violating Horava-Lifshitz fermionic term 
\begin{equation}\label{Lpsi}
{\cal L}_\psi = \bar\psi(i\slashed{D}_0+(i\slashed{D}_i)^3-m^3-\slashed{b}_0\gamma_5+(\slashed{b}\slashed{D}\slashed{D})_i\gamma_5)\psi,
\end{equation}
where $\slashed{D}_0=D_0\gamma^0$, $\slashed{D}_i=D_i\gamma^i$, $\slashed{b}_0=b_0\gamma^0$, and $(\slashed{b}\slashed{D}\slashed{D})_i=(bDD)_{ijk}\gamma^i\gamma^j\gamma^k$, with $D_{0,i}=\partial_{0,i}+ieA_{0,i}$ and, using the methodology applied in \cite{bumb2}, we define
\begin{equation}
(bDD)_{ijk}=\lambda_1b_iD_jD_k+\lambda_2b_jD_iD_k+\lambda_3b_kD_iD_j.
\end{equation}
We note that, since this Lagrangian involves only gauge covariant derivatives $D_{0,i}$, the quantum corrections will be gauge invariant.
In this Lagrangian, we have chosen the additive term linear in $b_{0,i}$ in order to break the CPT symmetry within $z=3$ theory.
It is more convenient to rewrite (\ref{Lpsi}) as
\begin{eqnarray}\label{Lpsi2}
{\cal L}_\psi &=& \bar\psi(i\slashed{\partial}_0+(i\slashed{\partial}_i)^3-m^3-\slashed{b}_0\gamma_5+(\slashed{b}\slashed{\partial}\slashed{\partial})_i\gamma_5-e\slashed{A}_0+e(\slashed{\partial}\slashed{\partial}\slashed{A})_i+ie(\slashed{b}\slashed{\partial}\slashed{A})_i\gamma_5 \nonumber\\
&&+ie^2(\slashed{\partial}\slashed{A}\slashed{A})_i-e^2(\slashed{b}\slashed{A}\slashed{A})_i\gamma_5-e^3\slashed{A}^3_i)\psi,
\end{eqnarray}
with
\begin{equation}
(b\partial\partial)_{ijk}=\lambda_1b_i\partial_j\partial_k+\lambda_2b_j\partial_i\partial_k+\lambda_3b_k\partial_i\partial_j,
\end{equation}
\begin{equation}
(\partial\partial A)_{ijk} = (\partial_i\partial_jA_k)+(\partial_jA_k)\partial_i+(\partial_iA_k)\partial_j+A_k\partial_i\partial_j+(\partial_iA_j)\partial_k+A_j\partial_i\partial_k+A_i\partial_j\partial_k,
\end{equation}
\begin{eqnarray}
(b\partial A)_{ijk} &=& \lambda_1b_i(\partial_jA_k)+\lambda_1b_iA_k\partial_j+\lambda_1b_iA_j\partial_k+ \lambda_2b_j(\partial_iA_k)+\lambda_2b_jA_k\partial_i+\lambda_2b_jA_i\partial_k \nonumber\\
&&+\lambda_3b_k(\partial_iA_j)+\lambda_3b_kA_j\partial_i+\lambda_3b_kA_i\partial_j,
\end{eqnarray}
\begin{equation}
(\partial AA)_{ijk} = (\partial_iA_j)A_k+A_j(\partial_iA_k)+A_jA_k\partial_i+A_i(\partial_jA_k)+A_iA_k\partial_j+A_iA_j\partial_k,
\end{equation}
and 
\begin{equation}
(bAA)_{ijk}=\lambda_1b_iA_jA_k+\lambda_2b_jA_iA_k+\lambda_3b_kA_iA_j.
\end{equation}

Then, the corresponding fermionic generating functional is
\begin{equation}
Z = \int D\bar\psi D\psi e^{i\int d^4x{\cal L}_\psi} = e^{iS_\mathrm{eff}},
\end{equation}
so that, by integrating out the spinor fields, we obtain the one-loop effective action  of the gauge field
\begin{eqnarray}\label{Seff}
S_\mathrm{eff} &=& -i\mathrm{Tr}\ln(\slashed{p}_0+\slashed{p}_ip_j^2-m^3-\slashed{b}_0\gamma_5-b_i\Lambda^i(p)\gamma_5-e\slashed{A}_0-e\Delta_i(k,p)A^i+eb_i\Lambda^{ij}(k,p)A_j\gamma_5 \nonumber\\
&&+e^2\nabla_{ij}(k,p)A^iA^j-e^2b_i\Lambda^{ijk}A_jA_k\gamma_5-e^3\slashed{A}_iA_j^2),
\end{eqnarray}
where
\begin{equation}
\Lambda^i(p)=\lambda_1 \gamma^i\slashed{p}_j\slashed{p}_k+\lambda_2\slashed{p}_j\gamma^i\slashed{p}_k+\lambda_3\slashed{p}_j\slashed{p}_k\gamma^i,
\end{equation}
\begin{equation}
\Delta^i(k,p) = \slashed{k}_j\slashed{k}_k\gamma^i+\slashed{p}_j\slashed{k}_k\gamma^i+\slashed{k}_j\slashed{p}_k\gamma^i+\slashed{p}_j\slashed{p}_k\gamma^i+\slashed{k}_j\gamma^i\slashed{p}_k+\slashed{p}_j\gamma^i\slashed{p}_k+\gamma^i\slashed{p}_j\slashed{p}_k,
\end{equation}
\begin{eqnarray}
\Lambda^{ij}(k,p) &=& \lambda_1\gamma^i\slashed{k}_k\gamma^j+\lambda_1\gamma^i\slashed{p}_k\gamma^j+\lambda_1\gamma^i\gamma^j\slashed{p}_k+\lambda_2\slashed{k}_k\gamma^i\gamma^j+\lambda_2\slashed{p}_k\gamma^i\gamma^j+\lambda_2\gamma^j\gamma^i\slashed{p}_k \nonumber\\
&&+\lambda_3\slashed{k}_k\gamma^j\gamma^i+\lambda_3\slashed{p}_k\gamma^j\gamma^i+\lambda_3\gamma^j\slashed{p}_k\gamma^i,
\end{eqnarray}
\begin{equation}
\nabla^{ij}(k,p) = \slashed{k}_k\gamma^i\gamma^j+\slashed{k}_k\gamma^j\gamma^i+\slashed{p}_k\gamma^i\gamma^j+\gamma^j\slashed{k}_k\gamma^i+\gamma^i\slashed{p}_k\gamma^j+\gamma^i\gamma^j\slashed{p}_k,
\end{equation}
and
\begin{equation}
\Lambda^{ijk} = \lambda_1\gamma^i\gamma^j\gamma^k+\lambda_2\gamma^j\gamma^i\gamma^k+\lambda_3\gamma^j\gamma^k\gamma^i,
\end{equation}
with $k_jA^i=i\partial_jA^i$. Here, $\mathrm{Tr}$ stands for the trace over the Dirac matrices, together with the trace over the integration in momentum and coordinate spaces.

In order to single out the quadratic terms in $A_\mu$ of the effective action,  we initially rewrite the expression (\ref{Seff}) as
\begin{equation}
S_\mathrm{eff}=S_\mathrm{eff}^{(0)}+\sum_{n=1}^\infty S_\mathrm{eff}^{(n)},
\end{equation}
where $S_\mathrm{eff}^{(0)}=-i\mathrm{Tr}\ln G^{-1}(p)$ and 
\begin{eqnarray}
S_\mathrm{eff}^{(n)} &=& \frac{i}{n}\mathrm{Tr}[G(p)(e\slashed{A}_0+e\Delta_i(k,p)A^i-eb_i\Lambda^{ij}(k,p)A_j\gamma_5 \nonumber\\
&&-e^2\nabla_{ij}(k,p)A^iA^j+e^2b_i\Lambda^{ijk}A_jA_k\gamma_5+e^3\slashed{A}_iA_j^2)]^n,
\end{eqnarray}
with $G(p)=(\slashed{p}_0+\slashed{p}_ip_j^2-m^3-\slashed{b}_0\gamma_5-b_i\Lambda^i(p)\gamma_5)^{-1}$.

Straightforward inspection shows that the graphs with one quartic or one quintic vertex simply cannot yield a desired CFJ-like structure. Hence the CFJ term is generated only by Feynman diagrams involving two triple vertices. So, let us calculate contributions from such diagrams. It should be noted that they superficially diverge, however, due to the gauge symmetry, divergent parts of these contributions actually vanish. In principle, this situation is analogous to that one in the usual Lorentz-breaking QED where the CFJ term, being also formally superficially divergent, is really finite and ambiguous, see f.e. \cite{ourrev}.

After evaluating the trace over the coordinate space, by using the commutation relation $A_\mu(x)G(p)=G(p-k)A_\mu(x)$ and the completeness relation of the momentum space, for the quadratic action $A_{0,i}$, we have
\begin{equation}\label{Seff1}
S_\mathrm{eff}^{(1)}\big|_{A^2} = i\int d^4x\, (\Pi_1^{ij}+\Pi_2^{ij})A_iA_j,
\end{equation}
with
\begin{equation}\label{Pi1}
\Pi_1^{ij} = -e^2\int\frac{d^{4}p}{(2\pi)^4}\mathrm{tr}\,G(p)(\nabla_{ij}(k,p)-b_m\Lambda^{mij}\gamma_5),
\end{equation}
and
\begin{equation}\label{Seff2}
S_\mathrm{eff}^{(2)}\big|_{A^2} = \frac i2 \int d^4x\, (\Pi_2^{00}A_0A_0+\Pi_3^{0j}A_0A_j+\Pi_4^{i0}A_iA_0+\Pi_5^{ij}A_iA_j),
\end{equation}
with
\begin{subequations}\label{Pi2}
\begin{eqnarray}
\label{Pi200}\Pi_2^{00} &=& e^2\int\frac{d^{4}p}{(2\pi)^4}\mathrm{tr}\,G(p)\gamma^0G(p-k)\gamma^0,\\
\label{Pi20j}\Pi_3^{0j} &=& e^2\int\frac{d^{4}p}{(2\pi)^4}\mathrm{tr}\,G(p)\gamma^0G(p-k)(\Delta^j(-k,p-k)-b_m\Lambda^{mj}(-k,p-k)\gamma_5), \\
\label{Pi2i0}\Pi_4^{i0} &=& e^2\int\frac{d^{4}p}{(2\pi)^4}\mathrm{tr}\,G(p)(\Delta^i(k,p)-b_m\Lambda^{mi}(k,p)\gamma_5)G(p-k)\gamma^0, \\
\label{Pi2ij}\Pi_5^{ij} &=& e^2\int\frac{d^{4}p}{(2\pi)^4}\mathrm{tr}\,G(p)(\Delta^i(k,p)-b_m\Lambda^{mi}(k,p)\gamma_5)G(p-k) \nonumber\\
&&\times(\Delta^j(-k,p-k)-b_n\Lambda^{nj}(-k,p-k)\gamma_5).
\end{eqnarray}
\end{subequations} 

Let us now single out the terms responsible for the generation of the CFJ term. For this, we must first consider the expansion 
\begin{equation}
G(p)=S(p)+S(p)(\slashed{b}_0\gamma_5+b_i\Lambda^i(p)\gamma_5)S(p)+\cdots,
\end{equation}
with $S(p)=(\slashed{p}_0+\slashed{p}_ip_j^2-m^3)^{-1}$, so that, initially, for (\ref{Pi1}), we have
\begin{equation}\label{Pi1a}
\Pi_1^{ij} = -e^2\int\frac{d^{4}p}{(2\pi)^4}\mathrm{tr}\,S(p)\slashed{b}_0\gamma_5S(p)\nabla_{ij}(k,p)+\cdots,
\end{equation}
where the dots mean higher orders in $b_{0,i}$ terms. From now on, we will omit these terms, as well as the higher derivative terms. To calculate the trace, we use first the simplest prescription for Dirac matrices, implying $\{\gamma^0,\gamma_5\}=0$ and $\{\gamma^i,\gamma_5\}=0$, so that we obtain
\begin{equation}
\Pi_1^{ij} = 4ie^2\int_{-\infty}^{\infty}\frac{dp_0}{2\pi}\mu^{3-d}\int\frac{d^{d}p}{(2\pi)^d}\frac{p_0^2-\vec p^6+m^6}{(p_0^2+\vec p^6+m^6)^2}\epsilon^{0ikj}b_0k_k, 
\end{equation}
where we have Wick rotated to Euclidean space ($p_0 \to ip_0$) and considered $d^3p/(2\pi)^3 \to \mu^{3-d}d^dp/(2\pi)^d$, with $\mu$ being an arbitrary scale parameter. Thus, after we calculate the integrals, we obtain
\begin{eqnarray}
\Pi_1^{ij} &=& -2e^2\mu^{3-d}\int\frac{d^{d}p}{(2\pi)^d}\frac{m^6}{(\vec p^6+m^6)^{3/2}}\epsilon^{0ikj}b_0k_k, \nonumber\\
&=& -\frac{e^22^{2-d} \pi ^{-\frac{d}{2}-\frac{1}{2}} m^{d-3} \mu^{3-d} \Gamma \left(\frac{3}{2}-\frac{d}{6}\right) \Gamma \left(\frac{d}{6}\right)}{3 \Gamma \left(\frac{d}{2}\right)}\epsilon^{0ikj}b_0k_k.
\end{eqnarray}

Let us now consider the expressions (\ref{Pi2}). The equation (\ref{Pi200}) does not contribute to CFJ term, and, after the calculation of the trace, (\ref{Pi20j}), (\ref{Pi2i0}), and (\ref{Pi2ij}) yield the results
\begin{eqnarray}
\Pi_3^{0j} &=& 4ie^2\int_{-\infty}^{\infty}\frac{dp_0}{2\pi}\mu^{3-d}\int\frac{d^{d}p}{(2\pi)^d}\frac{1}{(p_0^2+\vec p^6+m^6)^3}\epsilon^{i0kj}b_ik_k \nonumber\\
&&\times\frac1d[(\lambda_1+\lambda_3)\left(d m^{12}-(d+18) m^6 \vec p^6+(d+6) p_0^2 \vec p^6-d p_0^4+2 (d-3) \vec p^{12}\right) \nonumber\\
&&-\lambda_2\left(d m^{12}-(d+12) m^6 \vec p^6+(d-12) p_0^2 \vec p^6-d p_0^4+2 d \vec p^{12}\right)],
\end{eqnarray}
\begin{eqnarray}
\Pi_4^{i0} &=& 4ie^2\int_{-\infty}^{\infty}\frac{dp_0}{2\pi}\mu^{3-d}\int\frac{d^{d}p}{(2\pi)^d}\frac{1}{(p_0^2+\vec p^6+m^6)^3}\epsilon^{jik0}b_jk_k \nonumber\\
&&\times\frac1d[(\lambda_1+\lambda_3)\left(d m^{12}+3(2-3d) m^6 \vec p^6+(d+6) p_0^2 \vec p^6-d p_0^4+2 (d-3) \vec p^{12}\right) \nonumber\\
&&-\lambda_2\left(d m^{12}+3(4-3d) m^6 \vec p^6+(d-12) p_0^2 \vec p^6-d p_0^4+2 d \vec p^{12}\right)],
\end{eqnarray}
and
\begin{eqnarray}
\Pi_5^{ij} &=& 4ie^2\int_{-\infty}^{\infty}\frac{dp_0}{2\pi}\mu^{3-d}\int\frac{d^{d}p}{(2\pi)^d}\frac{\vec p^6}{(p_0^2+\vec p^6+m^6)^3} \nonumber\\
&&\times\{\epsilon^{ki0j}b_kk_0 \frac1d[(\lambda_1+\lambda_3)(-(3 d+16) m^6+d p_0^2+(d-4) \vec p^6) \\
&&-\lambda_2(-(3 d+14) m^6+(d-6) p_0^2+(d-2) \vec p^6)] + \epsilon^{0ikj}b_0k_k \frac3d(d+4) \left(-3 m^6-3 p_0^2+\vec p^6\right)\}, \nonumber
\end{eqnarray}
respectively. Now, when the integrals are calculated, we obtain
\begin{eqnarray}
\Pi_3^{0j} &=& -e^2\mu^{3-d}\int\frac{d^{d}p}{(2\pi)^d}\frac{\vec p^6}{(\vec p^6+m^6)^{5/2}}\epsilon^{i0kj}b_ik_k \left(\lambda _1-\lambda _2+\lambda _3\right) \frac1d \left((d-3) \vec p^6-2 (d+6) m^6\right), \nonumber\\
&=& \frac{e^22^{2-d} \pi ^{-\frac{d}{2}-\frac{1}{2}} \left(\lambda _1-\lambda _2+\lambda _3\right) m^{d-3}\mu^{3-d} \Gamma \left(\frac{3}{2}-\frac{d}{6}\right) \Gamma \left(\frac{d}{6}+2\right)}{\Gamma \left(\frac{d}{2}+1\right)}\epsilon^{i0kj}b_ik_k,
\end{eqnarray}
\begin{eqnarray}
\Pi_4^{i0} &=& -e^2\mu^{3-d}\int\frac{d^{d}p}{(2\pi)^d}\frac{\vec p^6}{(\vec p^6+m^6)^{5/2}}\epsilon^{jik0}b_jk_k \left(\lambda _1-\lambda _2+\lambda _3\right) \frac1d \left((d-3) \vec p^6+(6-8d) m^6\right), \nonumber\\
&=& \frac{e^22^{2-d} \pi ^{-\frac{d}{2}-\frac{1}{2}} \left(\lambda _1-\lambda _2+\lambda _3\right) m^{d-3}\mu^{3-d} \Gamma \left(\frac{3}{2}-\frac{d}{6}\right) \Gamma \left(\frac{d}{6}+1\right)}{\Gamma \left(\frac{d}{2}\right)}\epsilon^{jik0}b_jk_k,
\end{eqnarray}
and 
\begin{eqnarray}
\Pi_5^{ij} &=& -e^2\mu^{3-d}\int\frac{d^{d}p}{(2\pi)^d}\frac{\vec p^6}{(\vec p^6+m^6)^{5/2}}[\epsilon^{ki0j}b_kk_0 \left(\lambda _1-\lambda _2+\lambda _3\right) \frac1d \left((d-3) \vec p^6-2 (d+6) m^6\right) \nonumber\\
&&-\epsilon^{0ikj}b_0k_k \frac9d (d+4) m^6], \nonumber\\
&=& \frac{e^22^{2-d} \pi ^{-\frac{d}{2}-\frac{1}{2}} \left(\lambda _1-\lambda _2+\lambda _3\right) m^{d-3}\mu^{3-d} \Gamma \left(\frac{3}{2}-\frac{d}{6}\right) \Gamma \left(\frac{d}{6}+2\right)}{\Gamma \left(\frac{d}{2}+1\right)}\epsilon^{ki0j}b_kk_0 \nonumber\\
&&+\frac{e^22^{1-d} (d+4) \pi ^{-\frac{d}{2}-\frac{1}{2}} m^{d-3}\mu^{3-d} \Gamma \left(\frac{3}{2}-\frac{d}{6}\right) \Gamma \left(\frac{d}{6}\right)}{3 \Gamma \left(\frac{d}{2}\right)}\epsilon^{0ikj}b_0k_k. 
\end{eqnarray}

Therefore, with these results, by considering $S_\mathrm{eff}^{(1)}\big|_{A^2}+S_\mathrm{eff}^{(2)}\big|_{A^2} \to S_\mathrm{CFJ}$, we get the CFJ Lagrangian
\begin{eqnarray}
{\cal L}_{CFJ} &=& -\frac{2^{2-d} \pi ^{-\frac{d}{2}-\frac{1}{2}} m^{d-3}\mu^{3-d} \Gamma \left(\frac{3}{2}-\frac{d}{6}\right) \Gamma \left(\frac{d}{6}+1\right)}{\Gamma \left(\frac{d}{2}\right)} \{\left(\lambda _1-\lambda _2+\lambda _3\right) \\
&&\times[(\frac2d+\frac13) (b_i \epsilon^{i0jk}A_0\partial_j A_k+b_i \epsilon^{ij0k}A_j\partial_0 A_k)+b_i\epsilon^{ijk0}A_jk_kA_0]+b_0\epsilon^{0ijk}A_jk_jA_k\}.\nonumber
\end{eqnarray}
In order to ensure the gauge invariance, e.g., $\lambda_1=\lambda_2=\lambda_3=1$ and $d=3$, so that we obtain
\begin{equation}
{\cal L}_{CFJ} = -\frac{e^2}{4\pi^2} b_\kappa \epsilon^{\kappa\lambda\mu\nu}A_\lambda \partial_\mu A_\nu.
\end{equation}
We find that within this approach, the CFJ term is finite and reproduces one of the values obtained in the usual Lorentz-breaking QED \cite{ourrev}.

Now, let us use another prescription, that is, the 't Hooft and Veltman prescription for the calculations of the trace \cite{tHV} (for discussion of various regularization prescriptions within studies of the CFJ term, see f.e. \cite{TM}). Namely, we split the $d$-dimensional Dirac matrices $\gamma^i$ and the $d$-dimensional metric tensor $g^{ij}$ (with $\{\gamma^i,\gamma^j\}=2g^{ij}$ and $g_{ij}g^{ij}=d$) into $3$-dimensional parts and $(d − 3)$-dimensional parts, i.e., $\gamma^i=\bar\gamma^i+\hat\gamma^i$ and $g^{ij}=\bar g^{ij}+\hat g^{ij}$, so that now the Dirac matrices satisfy the relations
\begin{equation}\label{rel1}
\{\bar\gamma^i,\bar\gamma^j\}=2g^{ij},\ \{\hat\gamma^i,\hat\gamma^j\}=2g^{ij},\ \mathrm{and}\ \{\bar\gamma^i,\hat\gamma^j\}=0, 
\end{equation}
and, consequently, the metric tensors have the contractions 
\begin{equation}\label{cont}
\bar g_{ij}\bar g^{ij}=3,\ \hat g_{ij}\hat g^{ij}=d-3,\ \mathrm{and}\ \bar g_{ij}\hat g^{ij}=0. 
\end{equation}
However, the main change is into the relations 
\begin{equation}\label{rel2}
\{\bar\gamma^i,\gamma_5\}=0\ \mathrm{and}\ [\hat\gamma^i,\gamma_5]=0,
\end{equation}
where we have introduced the commutation relation of $\hat\gamma^i$ with $\gamma_5$. Note that the $\gamma^0$ matrix has the usual anticommutation relations, i.e.,
\begin{equation}\label{rel3}
\{\gamma^0,\gamma^0\}=2,\ \{\gamma^0,\gamma^i\}=0,\ \mathrm{and}\ \{\gamma^0,\gamma_5\}=0.
\end{equation}
Thus, by taking into account these relations (\ref{rel1}), (\ref{rel2}), and (\ref{rel3}), and contractions (\ref{cont}), after the calculation of the trace, the contribution (\ref{Pi1a}) becomes 
\begin{equation}
\Pi_1^{ij} = 4ie^2\int_{-\infty}^{\infty}\frac{dp_0}{2\pi}\mu^{3-d}\int\frac{d^{d}p}{(2\pi)^d}\frac{p_0^2+(1-\frac6d)\vec p^6+m^6}{(p_0^2+\vec p^6+m^6)^2}\epsilon^{0ikj}b_0k_k.
\end{equation}
Let us now calculate the integral which yield the result:
\begin{eqnarray}
\Pi_1^{ij} &=& -2e^2\mu^{3-d}\int\frac{d^{d}p}{(2\pi)^d}\frac{m^6+(1-\frac3d)\vec p^2}{(\vec p^6+m^6)^{3/2}}\epsilon^{0ikj}b_0k_k =%\nonumber\\&=& 
0.
\end{eqnarray}

The next step is to calculate the trace of the expressions (\ref{Pi2}). The results are 
\begin{equation}
\Pi_3^{0j}=\Pi_3\epsilon^{i0kj}b_ik_k,
\end{equation}
\begin{equation}
\Pi_4^{i0}=\Pi_3\epsilon^{jik0}b_jk_k,
\end{equation}
with
\begin{eqnarray}
\Pi_3 &=& 4ie^2\int_{-\infty}^{\infty}\frac{dp_0}{2\pi}\mu^{3-d}\int\frac{d^{d}p}{(2\pi)^d}\frac{1}{(p_0^2+\vec p^6+m^6)^3} \nonumber\\
&&\times\frac1d[(\lambda_1+\lambda_3)\left(d m^{12}-3 (d+4) m^6 \vec p^6+(7 d-12) p_0^2 \vec p^6-d p_0^4-4 (d-3) \vec p^{12}\right) \nonumber\\
&&-\lambda_2\left(d m^{12}-3 (d+2) m^6 \vec p^6+(7 d-30) p_0^2 \vec p^6-d p_0^4+2 (9-2 d) \vec p^{12}\right)],
\end{eqnarray}
and
\begin{eqnarray}
\Pi_5^{ij} &=& 4ie^2\int_{-\infty}^{\infty}\frac{dp_0}{2\pi}\mu^{3-d}\int\frac{d^{d}p}{(2\pi)^d}\frac{\vec p^6}{(p_0^2+\vec p^6+m^6)^3} \nonumber\\
&&\times\{\epsilon^{ki0j}b_kk_0 \frac1d[(\lambda_1+\lambda_3)(-(3 d+16) m^6+d p_0^2-(3 d-8) \vec p^6) \nonumber\\
&&-\lambda_2(-(3 d+14) m^6+(d-6) p_0^2-(3 d-10) \vec p^6)] \nonumber\\
&&- \epsilon^{0ikj}b_0k_k \frac7d \left((d+6)m^6+(d+6)p_0^2+(d-6)\vec p^6\right)\}.
\end{eqnarray}
Note that expression (\ref{Pi200}) does not contribute to CFJ term as well.

Now, by calculating the integrals, we can write 
\begin{equation}
\Pi_5^{ij}=\Pi_3(\epsilon^{ki0j}b_kk_0+\frac72 \left(\lambda _1-\lambda _2+\lambda _3\right)^{-1}\epsilon^{0ikj}b_0k_k),
\end{equation}
i.e., all the contributions (\ref{Pi2}), after the $p_0$ integral, are written in terms of $\Pi_3$, which has the form
\begin{eqnarray}
\Pi_3 &=& 2e^2\mu^{3-d}\int\frac{d^{d}p}{(2\pi)^d}\frac{\vec p^6}{(\vec p^6+m^6)^{5/2}} \left(\lambda _1-\lambda _2+\lambda _3\right) \frac1d \left((d-3) \vec p^6+(d+6) m^6\right) \nonumber\\
&=& 0.
\end{eqnarray}
Therefore,  by using the 't Hooft and Veltman prescription, one finds that $\Pi_1^{ij}=\Pi_2^{00}=\Pi_3^{0j}=\Pi_4^{i0}=\Pi_5^{ij}=0$ for the contributions to the CFJ term, so that
\begin{equation}
{\cal L}_{CFJ}=0, 
\end{equation}
i.e., we do not have the generation of the CFJ term within this prescription.  

Let us discuss our results. We studied the $z=3$ HL QED coupled to the constant axial vector $b_\mu$ and considered the contributions to the CFJ term. Just as in the usual Lorentz-breaking QED, this term was found to be superficially divergent but actually finite, with the result for it strongly depends on the calculation scheme, i.e., for the first time within studies of HL theories, we found a possibility to have a finite but ambiguous result. This result is rather natural since, as it was claimed in \cite{JackAmb}, the chiral anomaly is related with the ambiguity of the triangle graph, and since the existence of chiral anomalies in $z=3$ QED have been explicitly proved in \cite{Bakas}, the presence of ambiguity in the triangle graph in our theory seems to be rather natural, in a whole analogy with the usual Lorentz-breaking QED, cf. \cite{JackAmb}. Therefore, we note that our result confirms the conclusions obtained in \cite{Bakas}. Besides this, the possibility of the vanishing of the CFJ term in a certain calculation scheme can be treated as a reminiscence of the idea claimed in \cite{Altschul} that actually the most appropriate result for this term is zero.

{\bf Acknowledgements.} This work was partially supported by Conselho
Nacional de Desenvolvimento Cient\'{\i}fico e Tecnol\'{o}gico (CNPq).  The work by A. Yu. P. has been partially supported by the
CNPq project No. 301562/2019-9.


\begin{thebibliography}{99}
\bibitem{Lifshitz} E. M. Lifshitz, Zh. Eksp. Teor. Fiz., 11, 255 \& 269 (1941).
%\cite{Horava:2009uw}
\bibitem{Horava:2009uw}
P.~Horava,
%``Quantum Gravity at a Lifshitz Point,''
Phys.\ Rev.\ D \textbf{79}, 084008 (2009)
%doi:10.1103/PhysRevD.79.084008
[arXiv:0901.3775 [hep-th]].
\bibitem{EP}
C.~F.~Farias, J.~R.~Nascimento and A.~Y.~Petrov,
  %``On the effective potential for Horava-Lifshitz-like theories with the arbitrary critical exponent,''
  Phys.\ Lett.\ B {\bf 719}, 196 (2013)
%  doi:10.1016/j.physletb.2013.01.011
  [arXiv:1208.3427 [hep-th]].
\bibitem{CT}
M.~Gomes, T.~Mariz, J.~R.~Nascimento, A.~Y.~Petrov, J.~M.~Queiruga and A.~J.~da Silva,
  %``One-loop corrections in the Horava-Lifshitz-like QED,''
  Phys.\ Rev.\ D {\bf 92}, 065028 (2015);
  Erratum: [Phys.\ Rev.\ D {\bf 92}, 129902 (2015)]
%  doi:10.1103/PhysRevD.92.129902, 10.1103/PhysRevD.92.065028
  [arXiv:1504.04506 [hep-th]].
\bibitem{Lima} A.~M.~Lima, T.~Mariz, R.~Martinez, J.~R.~Nascimento, A.~Y.~Petrov and R.~F.~Ribeiro,
  %``Horava-Lifshitz-like Gross-Neveu model,''
  Phys.\ Rev.\ D {\bf 95}, 065031 (2017)
%  doi:10.1103/PhysRevD.95.065031
  [arXiv:1612.05900 [hep-th]];
  M.~Gomes, T.~Mariz, J.~R.~Nascimento, A.~Y.~Petrov and A.~J.~da Silva,
  %``1/N Expansion for Horava-Lifshitz like four-fermion models,''
   Eur. Phys. J. C {\bf 80} (2020), 518
  [arXiv:2001.06467 [hep-th]].
\bibitem{Ruben}   T.~Mariz, R.~Moreira and A.~Y.~Petrov,
  %``Emergent gauge bosons and dynamical symmetry breaking in a four-fermion Lifshitz model,''
  Eur.\ Phys.\ J.\ C {\bf 79}, 550 (2019)
  Erratum: [Eur.\ Phys.\ J.\ C {\bf 79}, 729 (2019)]
 % doi:10.1140/epjc/s10052-019-7068-x, 10.1140/epjc/s10052-019-7243-0
  [arXiv:1905.04130 [hep-th]].
\bibitem{JK} %\bibitem{Jackiw:1999yp}
R.~Jackiw and V.~Kostelecky,
%``Radiatively induced Lorentz and CPT violation in electrodynamics,''
Phys. Rev. Lett. \textbf{82} (1999), 3572-3575
%doi:10.1103/PhysRevLett.82.3572
[arXiv:hep-ph/9901358 [hep-ph]].  
\bibitem{ourrev} A.~Ferrari, J.~R. Nascimento and A.~Y.~Petrov,
%``Radiative corrections and Lorentz violation,''
Eur. Phys. J. C \textbf{80} (2020), 459
%doi:10.1140/epjc/s10052-020-8000-0
[arXiv:1812.01702 [hep-th]].
\bibitem{TMP} T.~Mariz, J.~R.~Nascimento and A.~Y.~Petrov,
  %``On the Adler-Bell-Jackiw anomaly in a Horava-Lifshitz?like QED,''
  EPL {\bf 112}, 61002 (2015)
 % doi:10.1209/0295-5075/112/61002
  [arXiv:1505.00715 [hep-th]].
\bibitem{bumb2} T.~Mariz, J.~R. Nascimento and A.~Y.~Petrov,
%``Horava-Lifshitz four-fermion model revisited and dynamical symmetry breaking,''
Phys. Rev. D \textbf{101} (2020), 105008
%doi:10.1103/PhysRevD.101.105008
[arXiv:1912.13378 [hep-th]].
\bibitem{tHV} G.~'t Hooft and M.~J.~G.~Veltman,
  %``Regularization and Renormalization of Gauge Fields,''
  Nucl.\ Phys.\ B {\bf 44}, 189 (1972).
\bibitem{TM} T.~Mariz, J.~R.~Nascimento, E.~Passos, R.~F.~Ribeiro and F.~A.~Brito,
  %``A Remark on Lorentz violation at finite temperature,''
  JHEP {\bf 0510}, 019 (2005)
  %doi:10.1088/1126-6708/2005/10/019
  [hep-th/0509008]; 
  T.~Mariz, J.~R.~Nascimento and A.~Y.~Petrov,
  %``On the perturbative generation of the higher-derivative Lorentz-breaking terms,''
  Phys.\ Rev.\ D {\bf 85}, 125003 (2012)
%  doi:10.1103/PhysRevD.85.125003
  [arXiv:1111.0198 [hep-th]];
  J.~F.~Assun\c{c}\~{a}o and T.~Mariz,
  %``Radiatively induced CPT-odd Chern-Simons term in massless QED,''
  EPL {\bf 110}, 41002 (2015)
%  doi:10.1209/0295-5075/110/41002
  [arXiv:1505.08156 [hep-th]].
\bibitem{JackAmb} R.~Jackiw,
%``When radiative corrections are finite but undetermined,''
Int. J. Mod. Phys. B \textbf{14} (2000), 2011-2022
%doi:10.1142/S021797920000114X
[arXiv:hep-th/9903044 [hep-th]].
\bibitem{Bakas} I.~Bakas and D.~Lust,
%``Axial anomalies of Lifshitz fermions,''
Fortsch. Phys. \textbf{59} (2011), 937
%doi:10.1002/prop.201100048
[arXiv:1103.5693 [hep-th]];
I.~Bakas,
%``More on axial anomalies of Lifshitz fermions,''
Fortsch. Phys. \textbf{60} (2012), 224-242
%doi:10.1002/prop.201100086
[arXiv:1110.1332 [hep-th]].
\bibitem{Altschul} B.~Altschul,
%``There is No Ambiguity in the Radiatively Induced Gravitational Chern-Simons Term,''
Phys. Rev. D \textbf{99} (2019), 125009
%doi:10.1103/PhysRevD.99.125009
[arXiv:1903.10100 [hep-th]].

\end{thebibliography}
\end{document}